\documentclass[11pt]{article}    
\usepackage{psfig}

\begin{document}  
 
\newcommand{\bra}{\langle}     
\newcommand{\ket}{\rangle}       
\newcommand{\beq}{\begin{equation}}       
\newcommand{\eeq}{\end{equation}}       
\def\bea{\begin{eqnarray}}
\def\eea{\end{eqnarray}}
\def\tr{{\rm tr}\,} 
\def\href#1#2{#2}

\begin{titlepage} 
 
\begin{center} 
\hfill UW/PT-02/10 \\ 
\hfill hep-th/0205236 
\vspace{2cm} 
 
{\Large\bf Adding flavor to AdS/CFT}

\vspace{1.5cm} 
{\large 
Andreas Karch and Emanuel Katz } 
 
\vspace{.7cm}

{Department of Physics,
University of Washington,
Seattle, WA 98195, USA\\
[.4cm]}
({\tt karch@phys.washington.edu}) \\
({\tt amikatz@phys.washington.edu})
 
\end{center} 
\vspace{1.5cm} 
 

Coupling fundamental quarks to QCD in the dual string representation 
corresponds to adding the open string sector. Flavors 
therefore should be represented by space-time filling D-branes in the dual 5d
closed string background. 
This requires several interesting properties of D-branes in AdS.
D-branes have to be able to end in thin air
in order to account for massive quarks, which only live in the UV region.
They must come in distinct sets, representing the chiral global
symmetry, with a bifundamental field playing the role
of the chiral condensate.
We show that these expectations are born out in several
supersymmetric examples.
To analyze most of these properties it is not necessary to go
beyond the probe limit in which one neglects the
backreaction of the flavor D-branes.
\vspace{3.5cm} 
\begin{center} 
\today 
\end{center} 
\end{titlepage} 

\section{Introduction}

In the large $N$ limit the Feynman diagrams of $SU(N)$ Yang-Mills theory
reorganize themselves into a genus expansion of closed string 
theory \cite{hooft}.
This closed string is believed to propagate in a five dimensional
background \cite{polyakov}. 
At each point on the worldvolume of the string, we have
to specify its position in 4d Minkowski space and its thickness, which
is represented by its position in the 5th dimension. The metric
structure of the 5th dimension describes the internal structure of the string.
This expectation has been realized in many supersymmetric examples 
starting with \cite{juan},
where in addition to the 5th dimension, the closed string background includes
an internal compact space, representing the additional fields in the
theory. Thus, pure Yang-Mills, with only glue as degrees of freedom, 
is expected to map to an entirely 5d non-critical string theory 
background \cite{polyakov}.

Adding fundamental flavors effectively introduces boundaries in the 't Hooft
expansion, that is one adds an open string sector. 
Since the open strings should be allowed to have a thickness
as well, one is led to believe that adding fundamental flavors in the
gauge theory maps to adding space-time filling D-branes in the 5d bulk
theory.  In the limit where $M$, the number of flavors, is much
smaller than $N$, the backreaction of the D-branes on the
bulk geometry can be
ignored. As we discuss later, this corresponds
to the quenched approximation of lattice gauge theory. Several
puzzles arise. The spacetime filling D-branes can't carry
any charge in order to avoid tadpoles, but nevertheless should be
stable. Making the quarks very heavy should decouple them
from the IR theory, so D-branes dual to massive
quarks should be spacetime filling in the UV region but than end
at a finite distance in the 5th dimension, and be absent in the IR.
The gauge fields living on the D-branes
map to global flavor currents in the gauge theory. 
Since QCD with $M$ flavors has a chiral $SU(M) \times SU(M)$ symmetry,
which then gets broken by a condensate, one actually needs
two sets of space-time filling branes each carrying
one of the $SU(M)$ symmetries.  The chiral condensate should then 
manifest itself in the bulk as a vev for a scalar, 
which is bifundamental under the product gauge symmetry. 
We will show that all this is indeed 
the case in the supersymmetric cousins of QCD,
where the dual closed string background is known.

In the next section we will consider the probe limit,
and show how to introduce space-time filling branes without
running into contradictions with tadpoles. Most of the stable D-branes
we know carry charge under RR-gauge fields, so that without
introducing orientifolds it is usually impossible to introduce
space-time filling branes, since the net charge has to cancel.
We avoid this problem by having our D-branes wrap topologically
trivial cycles with zero flux, so that they carry no charge in the
5d world. They are stabilized by the same mechanism as in \cite{kr2,bp}:
there are negative mass modes which control the slipping of the D-branes off
the cycle they wrap, but the mass is not negative enough to lead
to an instability in the curved 5d geometry, e.g. for our conformal
examples it is above the BF bound \cite{BF,BF2}.
We show how to read off both the cycle wrapped
in the internal manifold, and the worldvolume
scalars that are turned on, from knowledge
of the embedding in the flat 10d spacetime. Giving
a mass to the flavors turns on a non-trivial profile
for the slipping mode, making the D-brane seemingly terminate from
the 5d point of view. 
Last but not least,
we will study in this section a simple model that has
an $SU(M) \times SU(M)$ global symmetry and show how this
is reproduced in the bulk by having two distinct spacetime filling
sets of branes, which differ by the cycle they wrap in the internal 
space. We explicitly break the product global symmetry to the diagonal by 
turning on a mass for the quarks.  In the corresponding bulk physics this 
amounts to Higgsing the gauge symmetry by a bifundamental scalar, mimicking 
the bulk physics dual to a chiral condensate.
We will also verify that all our solutions are
supersymmetric by analyzing $\kappa$-symmetry on the worldvolume.

Section 3 will be devoted to our most interesting example,
a close cousin of ${\cal N}=1$ SYM with $M$ flavors. This
theory actually has a chiral global symmetry.
Again we will show how to obtain the $SU(M) \times SU(M)$ symmetry
in the bulk from two sets of D-branes, but this time
a single D7 brane will only give rise to a single chiral multiplet. 
As opposed to the examples studied before, the ${\cal N}=1$ theory has
non-trivial $B$ field turned on in the supergravity background. So this
time a lower dimensional D-brane charge is induced and tadpole
cancellation becomes a non-trivial constraint, corresponding
to the dual gauge theory being anomaly free.
In Section 4 we conclude and speculate about the use of D-brane 
probes to extract the non-perturbative superpotential of SQCD.

\section{Spacetime filling branes and Tadpoles}
\subsection{The probe limit}
In this work we only study branes in the probe limit, as 
in \cite{kr2}.
This is justified by taking the limit where $g_s \rightarrow 0$,
$M$ fixed, such that $g_s M \rightarrow 0$, which is the strength with which
the $M$ D-branes source the metric, dilaton and gauge fields. 
At the same time we take $N \rightarrow \infty$ holding as usual
$g_s N = \lambda$ fixed. So the D3 brane backreaction is large, and we 
replace the D3 branes with their near-horizon geometry. The flavor probe
branes will minimize their worldvolume action in this background without
deforming the background. In particular this means that we take
$M<<N$, that is in the large $N$ gauge theory we introduce a finite number
of flavors. In the lattice literature this limit is known as the quenched
approximation: the full dynamics of the glue and its effect on the fermions
is included, but the backreaction of the fermions on the glue is dropped.
In the probe limit this approximation becomes exact. For some of the
backgrounds we study, like the D3-D7 system, the supergravity solution
is known beyond the probe limit \cite{AFM,grana,italy1,italy2}.

\subsection{Tadpoles}

As explained in the introduction, our proposal that fundamental quarks
are dual to spacetime filling D-branes seems to suffer from a
problem with tadpoles: usually stable D-branes carry charge, and 
for a spacetime filling brane, the net charge has to cancel. 
The way the supersymmetric
branes we study avoid this pitfall is that they
wrap topologically trivial cycles in the internal space.
This phenomenon was encountered in \cite{kr2} for the closely
related D3-D5 system. There the field theory interpretation
is in terms of a CFT with defect or boundary (dCFT). 
The $M$ D5 branes give rise to $M$ flavors
in the dual gauge theory, but they are confined to a codimension
one defect. In the supergravity dual, the branes aren't spacetime
filling but live on an AdS$_4$ inside the AdS$_5$. For the
cosmological constant to not jump across the branes, they also should
not carry any D3 brane charge. This happens since the D5 wraps
an equatorial, topologically trivial $S^2$ inside the $S^5$. Also
the $S^2$ is not stabilized by any flux, so there is no induced
D3 brane charge. The brane is stabilized 
by the dynamics of fields in AdS. As shown by Breitenlohner and Freedman
\cite{BF},
a negative mass mode in AdS$_{d+1}$ does not lead to an instability as long
as the mass is above the BF bound, $m^2 \geq - \frac{d^2}{4}$ in units
of the curvature radius. Roughly speaking, this is due to AdS being a box.
To satisfy the boundary conditions, all non-vanishing fluctuations
have to carry a finite amount of gradient energy. In order to lower the
energy of the system by rolling down the hill, the field's 
loss in potential energy would have to outweigh its gain in gradient energy. 
As shown in \cite{kr2} the
slipping mode that wants to have the D5 move off the equator and contract
to a point, as allowed by topology, is such a stable,
negative mass mode above the BF
bound. 

The same mechanism can be employed to get flavors
from spacetime filling D-branes.
The simplest example 
is the D3-D7 system. The corresponding
gauge theory is ${\cal N}=4$ $SU(N)$ SYM with $M$
additional fundamental ${\cal N}=2$ hypermultiplets. This
theory is asymptotically non-free, but the $\beta$-function
is proportional to $g^2_{YM} M$, so that in the probe
limit the theory is still conformal, the background is
just the usual $AdS_5 \times S^5$. Writing the $S^5$ as
\beq
\label{dd}
ds^2 = L^2 (d \psi^2 + \cos^2(\psi) d \theta^2 + \sin^2(\psi) d \Omega_3^2)
\eeq
the D7 branes wrap an
$S^3$ given by
\beq
\psi=0
\eeq
which can easily be seen from its flat space embedding as we will 
explain below.
Expanding the D7 brane action to quadratic order in the fluctuations of 
$Z_{\psi}$ around the extremum $\psi=\frac{\pi}{2}$ (as
was done in \cite{DFO} for
the D3 D5 system)
\beq
{\cal L} \sim 1 + \frac{1}{2} (\partial Z_{\psi})^2 - \frac{3}{2} Z_{\psi}^2,
\eeq
we can read of the
masses of scalar fluctuations and the corresponding operator dimensions
\beq
m_l^2=l (l+2) -3, \; \; \; \Delta_l = 1-l, 3+l.
\eeq
The dimension 3, $l=0$, mode maps to the fermion mass term in the CFT.
This spectrum was already anticipated in \cite{AFM} without
actually doing the calculation. It was
noted that due to superconformal invariance, the r-charge and
the dimension of the chiral operators are related, so that one can
simply determine the dimension of a given spherical harmonic on the $S^3$
from its angular quantum numbers.

In order to establish that this $S^3$ is actually a supersymmetric solution to
the worldvolume equations of motion, we have to check $\kappa$ symmetry
on the worldvolume. We will do so for this and similar configurations
at the end of this section.

\subsection{Worldvolume Scalars: D-branes vanishing in thin air}

In order to determine the above cycle on which 
the 7-branes wrap the internal
manifold, we used our knowledge of the brane
setup in the full asymptotic space-time. The coordinate
\beq
u^2 =  x_4^2 +x_5^2 +x_6^2 +x_7^2 +x_8^2 +x_9^2 
\eeq
becomes the radial coordinate in AdS$_5$ and the $S^5$ is
the $u=const.$ slice of the $R^6$ parametrized by the 4,5,6,7,8,9
coordinates.
The D7 sitting on
\beq x_8=x_9=0
\eeq
wraps an equatorial $S^3$ inside the $S^5$ and fills all of AdS$_5$.
Generically the cycle wrapped by the D7 brane will not have this simple
product form, but the cycle in the internal space will be fibered over
the AdS space. From the 5-dimensional point of view this is encoded
in a non-trivial profile for one of the D-brane worldvolume scalars.
To see how this works let us study the D3-D5 system of \cite{kr2}
with a non-zero mass term for the defect hyper. In \cite{kr2}
similar logic was used to establish that a D5 brane along $x_3=x_7=x_8=x_9=0$,
wraps the equatorial $S^2$ inside the $S^5$.
In the AdS$_5$, whose metric is given by
\beq
ds^2 = L^2 \left ( \frac{du^2}{u^2} + u^2 (-dx_0^2 + dx_1^2 + dx_2^2 + dx_3^2)
\right ),
\eeq 
it lives on $x_3=0$; that is on an AdS$_4$ slice with
the same curvature radius, stretching straight
from the boundary towards the horizon. The field theory
described by this brane system is 4d ${\cal N}=4$ SYM  
with a 3d hypermultiplet defect \cite{HW,kr2,DFO}.

The simple product form of the worldvolume metric gets modified when
turning on a mass term for the defect hyper, by having the D5-brane
now at $x_7=c$. Using a parametrization of the $S^5$ where 
$x_7 = u \cos (\psi)$ this leads to \beq
\label{psi}
\psi =
\arccos(\frac{c}{u}) \sim \frac{\pi}{2} - \frac{c}{u} - \frac{1}{6} \frac{c^3}{u^3} +
\ldots, 
\eeq
where on the D5 worldvolume $u$ only takes values between $\infty$ and $c$.
For $c=0$ we recover the equatorial $S^2$ sitting at $\psi=\frac{\pi}{2}$.
Since $\psi$ is one of the worldvolume scalars, we see that
non-zero $c$ induces a $u$-dependent profile for this scalar.
The tension of the D5, that is the 
potential for $\psi$, is proportional to the radius of the $S^2$,
 $\cos^2 (\psi)$. At $u=c$ the tension of the D-brane
becomes zero and the D5 brane just stops. This scalar profile
can be interpreted as an exactly solvable boundary RG 
flow\footnote{Recently a different boundary RG flow appeared
in \cite{skenderis}. Their flow breaks all supersymmetry and
corresponds to adding an operator of dimension 4. There, the actual
embedding of the brane changes, and only asymptotically does it
become AdS$_4$ $\times$ $S^2$.}.

This at first sounds implausible, D-branes should not be able to
end in the middle of nowhere. In our case, this is possible
due to the existence of the BF bound in AdS: for one we have
a stable D-brane without charge, so that the usual reasoning that
the D-brane has to end on something which carries away its flux, fails.
The fact that the tension can go to zero at a point in space relies
on the fact, that we can have a scalar which lowers the tension
without leading to an instability.
From the higher dimensional point of view it is clear that
the D-brane isn't really ending at all since the $S^2$ goes to
zero size. Similar phenomena have appeared 
in the literature, e.g. in \cite{AV} where an internal circle
shrinks when a D6 brane "ends". What is different there
is that spacetime itself is a circle fibration. Compactifying
on the internal space, the locus of the degenerate fiber becomes
a defect, on which the D-brane then ends. In our case the
background is a product and the $S^5$ has constant size. It is only
the $S^2$ the brane wraps that vanishes. In the lower dimensional
description the brane literally vanishes in thin air.

From the dual dCFT point of view, it is also clear what is going on.
The leading $\frac{c}{u}$ term means we deformed by the dimension
2 operator dual to $\psi$, so that the boundary behavior is like 
$\frac{1}{u}^{d- \Delta}=\frac{1}{u}$.
This is just the mass term for the defect hyper.
The absence of the subleading $\frac{1}{u^2}$ term
indicates that no vev
is turned on. In the dCFT the defect has to become transparent at energies
below the mass of the defect hyper. It is amazing to see how directly
the gravity captures this phenomenon: the D-brane stops at the
energy scale set by the mass of the defect hyper. 

\begin{figure}
 \centerline{\psfig{figure=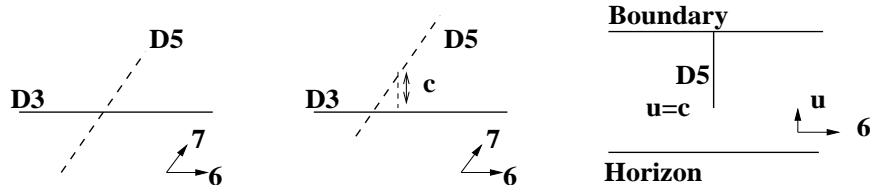,width=4.5in}}
 \caption{D3 D5 system with mass term for defect scalars,
in the corresponding 5d AdS spacetime, the D5 brane ends in the middle
of nowhere.}
\label{thinair}
  \end{figure}

\subsection{Interactions and the subleading term in AdS/CFT}

The behavior of the D3-D5 defect system
generalizes to the case of large $N$ gauge theory
with massive flavors, e.g. the D3-D7 system, with
the D7 brane separated from the D3 branes in $x_8$. 
The D-brane representing the massive fundamentals is spacetime filling
in the UV and then simply stops at the position dual to the mass
scale. The exact profile for $\psi$ is once more known and
given by (\ref{psi}). The leading
term again corresponds to the fermion mass term, this time
of dimension 3. However, here, the subleading order $u^{-3}$ term seems to
indicate the presence of a vev. We know from the field theory that this
is not the case. We will show that this is an artifact
of treating the scalar as a free field.
According to the rules of AdS/CFT \cite{gubser,witten},
we have to evaluate the classical action on the solution.
In the presence of a leading $c/u$ term higher order
terms from the non-linearities in the action, quartic in
the leading term, come in at the same order as the first contribution
of the subleading $c^3/u^3$ term. We use a coordinate
$r=\log(u)$ in terms of which the metric reads 
\beq ds^2 = e^{2r} (-dt^2 + d\vec{x}^2) + dr^2 \eeq
In terms of the fluctuating field, defined via
$\psi= \frac{\pi}{2} + \varphi(r)$,
the DBI action expanded to quartic order in $\varphi$ reads
\beq
 S= \int d^5x e^{4r} \sin^3(\psi) \sqrt{1 + (\varphi')^2} =
(1-\frac{3}{2} \varphi^2 + \frac{7}{8} \varphi^4+ \ldots )(1+
\frac{1}{2} (\varphi')^2 - \frac{1}{8} (\varphi')^4 + \ldots).
\eeq
Evaluated on 
\beq
\varphi =  - c\, e^{-r} - \frac{1}{6} c^3 e^{-3r},
\eeq
\beq
{\cal L} = 1- (c \, e^{-r})^2
\eeq
exactly.  All higher order terms cancel on our solution.
No vacuum expectation value is thus turned on.
A very similar result was obtained in \cite{bianchi1,bianchi2}
where it was also found that the definition of the subleading term in the
presence of the leading term requires a more careful treatment.

\subsection{$SU(M_1) \times SU(M_2)$ global symmetry}

The third property we would like the spacetime filling D-branes to have,
is to come in two distinct sets, giving rise to chiral $SU(M) \times
SU(M)$ global symmetry, as in QCD. Before we study a chiral
example in the next section, let us exhibit this effect
in a simpler system, which has a non-chiral $SU(M_1) \times SU(M_2)$ global
symmetry, the 
D3-D7-D7' system. The D7 and D7' branes are distinguished by their embedding
in the 6 directions transverse to the D3, parametrized by 3 complex 
variables $u_1$, $u_2$,
$u_3$. The D7 branes we studied so far sit at $u_3 = x_8 + i \; x_9=0$.
D7' branes sit at $u_2=0$, breaking another half of the supersymmetries.

The worldvolume gauge theory is ${\cal N }=4$ SYM with $M_1 + M_2$ extra
hypermultiplets $Q$ and $T$. 
Each of the flavors forms a different ${\cal N}=2$
subsystem, that is it couples to a different adjoint \cite{yin}:
\beq
W= X [Y,Z] + Y Q \tilde{Q} + Z T \tilde{T}. 
\eeq
This way only ${\cal N}=1$ supersymmetry is preserved, and the global
symmetry is $SU(M_1) \times SU(M_2)$.
Writing the metric on $S^5$ as the metric inherited from the $C^3$
formed out of the 3 $u_i = r_i e^{i \theta_i}$ subject to
$\sum r_i^2 =1$,
the two D7 branes are given by the two $S^3$s $u_2$=0 and
$u_3=0$ respectively. 
Since they intersect along $u_2=u_3=0$ in $C^3$, on the $S^5$ they
intersect over a circle. They give rise to two separate spacetime
filling 4-branes in AdS$_5$, responsible for the two global
symmetries. The reduction of the 8d fields on the equatorial $S^3$s
proceeds as before. On the intersection, the 77' strings
give rise to a 6d, bifundamental, hypermultiplet. It's KK reduction on the $S^1$
gives rise to scalar fields dual to operators mixing $Q$ and $T$ flavors.

From the bulk gravity point of view, 
a toy model for chiral symmetry breaking is the mass
deformation
\beq
\Delta W =h Q \tilde{T} + \tilde{h} \tilde{Q} T
\eeq
This corresponds to giving a vev to the 6d bifundamental hyper \cite{yin}.
For $M_1=M_2=M$, the global
$SU(M) \times SU(M)$ symmetry in the field theory
gets broken explicitly to the
diagonal $SU(M)$. The 2 disjoint 7 branes join into one smooth
7-brane wrapping the curve \cite{mina1, mina2} 
\beq
\label{emb}
u_2 u_3 = \epsilon \sim h \tilde{h}.
\eeq
From the bulk point of view, the gauge symmetry gets higgsed to
the diagonal subgroup. The scalars in the 6d hypermultiplet develop
a non-trivial profile.
This is precisely the way chiral symmetry breaking would be
captured by the bulk physics, the only difference being that
for chiral symmetry breaking only subleading terms would be present
in the profile.
A geometric measure for the vev $h \tilde{h}$
is the size
of the disk ending on the curve \cite{mina1}. The curve
$u_2 u_3 = \epsilon$ in $R^6$ has a non-contractible circle of radius
$R=\sqrt{2 \epsilon}$. As we will show in the appendix, on the $S^5$,
the corresponding curve still has a non-trivial circle. Its radius
at a given value of $u$ is given by $R=\frac{\sqrt{2 \epsilon}}{u}$
consistent with having a fermion mass term turned on. To see the absence
of a sub-leading term, which would be dual to a fermion-bilinear vev, 
we would need a better understanding of the non-linear completion
of the 6d HM action, that is, the analog of the DBI used previously.

\subsection{$\kappa$- Symmetry}

To
get a BPS brane in spacetime, supersymmetry variations of the
background fields have to be compensated by a $\kappa$ symmetry
transformation on the worldvolume of the D7 branes. For this to work, one
needs to satisfy \cite{cederwall1,cederwall2,townsend,kallosh,bilal}
\beq \Gamma \epsilon = \epsilon
\eeq
where
\beq \Gamma= e^{-\frac{a}{2}} \Gamma'_0 e^{\frac{a}{2}}
\eeq
with \beq
\Gamma'_0 = i \sigma_2 \otimes \frac{1}{(8)! \sqrt{g} }
\epsilon^{i_1 \ldots i_{8}} \partial_{i_1} X^{a_1} \cdot \ldots
\cdot \partial_{i_{8}} X^{a_{8}} \; \Gamma'_{a_1 \ldots a_{8}},
\eeq
the matrix $a$ captures the $B$-field and worldvolume gauge field, in our case
$B=0$, $a=1$. $\Gamma'$s are curved space $\Gamma$-matrices, 
$\Gamma'_a = E_a^M \Gamma_M$,
and $\epsilon$ is a Killing spinor of the AdS$_5$ $\times$ $S^5$
background, whose explicit form can found in 
\cite{myers,rahmfeld}. For the metric eq.(\ref{dd}) we used
in the D3-D7 system:
\beq
\epsilon = 
e^{\frac{i}{2} \psi \gamma_5 \Gamma^{\psi} }
e^{\frac{i}{2} \theta \gamma_5 \Gamma^{\theta} }
e^{-\frac{i}{2} \alpha_1 \Gamma^{\alpha_1 \psi} }
e^{-\frac{i}{2} \alpha_2 \Gamma^{\alpha_2 \alpha_1} }
e^{-\frac{i}{2} \alpha_3 \Gamma^{\alpha_3 \alpha_2} } 
 \times R_{AdS} \epsilon_0
\eeq
where $\gamma_5 = \Gamma^{\psi} \Gamma^{\theta} \Gamma^{\vec{\alpha}}$, 
$\Gamma^{\vec{\alpha}}=
\Gamma^{\alpha_1 \alpha_2 \alpha_3}$,
$R_{AdS}$ is a similar rotation matrix to the one we spelled out for
the sphere \cite{myers,rahmfeld} and
$\epsilon_0$ is a constant spinor. What is important is that all the sphere 
$\Gamma$ matrices
as well as $\gamma$, the analog of $\gamma_5$ for the 5 AdS directions, 
commute with $R_{AdS}$.

The projection matrix for the D7 living on the equator,
$\psi=\frac{\pi}{2}$, simply becomes
\beq
\Gamma= -\Gamma^{\vec{\alpha}} \gamma.
\eeq
Now we want to solve $\Gamma \epsilon=\epsilon$. First, note that since all the
exponentials contain an even number of $\Gamma$ matrices associated with the
sphere, $\gamma$ pulls through all the exponentials and can be made to act
on $\epsilon_0$. If we try to do the same with $\Gamma^{\vec{\alpha}}$, we
flip the sign of the first exponentials on the left hand side (these are the
terms with an odd number of $\Gamma$ matrices differing from the one
we want to pull through). Multiplying both sides with $e^{\frac{-i}{2} \psi
\gamma_5 \Gamma^{\psi} }$ we arrive at
\begin{eqnarray}
\nonumber
-e^{-i \psi \gamma_5 \Gamma^{\psi}} 
e^{\frac{-i}{2} \theta \gamma_5 \Gamma^{\theta}} 
e^{\frac{i}{2} \alpha_1 \Gamma^{\alpha_1 \psi} }
e^{-\frac{i}{2} \alpha_2 \Gamma^{\alpha_2 \alpha_1} }
e^{-\frac{i}{2} \alpha_3 \Gamma^{\alpha_3 \alpha_2} } 
 \times R_{AdS} \Gamma^{\vec{\alpha}} \gamma  \epsilon_0 &=& \\
e^{\frac{i}{2} \theta \gamma_5 \Gamma^{\theta}} 
e^{-\frac{i}{2} \alpha_1 \Gamma^{\alpha_1 \psi} }
e^{-\frac{i}{2} \alpha_2 \Gamma^{\alpha_2 \alpha_1} }
e^{-\frac{i}{2} \alpha_3 \Gamma^{\alpha_3 \alpha_2} }
\times R_{AdS}  \epsilon_0.&&
\end{eqnarray}
At $\psi=\frac{\pi}{2}$, the first term on the lhs just becomes 
$-i \gamma_5 \Gamma^{\psi}$.
Pulling this 
through changes the next two exponentials back to agree with the rhs
and we arrive at a simple projector acting on $\epsilon_0$.
The D7 brane worldvolume theory hence preserves 16 out of the
32 supercharges of the AdS$_5$ $\times$ $S^5$ background. In the field
theory this reflects the fact that, in the probe limit, the theory
with massless flavors still preserves conformal invariance and
hence the resulting ${\cal N}=2$ theory actually has 16 instead
of 8 supercharges. 

Next, let us consider what happens when we turn on the mass parameter, that is 
we separate the D7 brane from the D3 branes. The superconformal generators
now will be broken, but the D7 brane still preserves 8 supercharges.
This actually follows from a very general result of Kehagias
\cite{kehagias}, as discussed in \cite{grana2}.
In a type IIB background of the
form
\beq
ds^2 = Z^{-\frac{1}{2}} \eta_{\mu \nu} + Z^{\frac{1}{2}} ds_K^2
\eeq
where the warpfactor $Z$ depends on the internal Ricci flat space $K$
\beq \epsilon = Z^{-\frac{1}{8}} \epsilon_0
\eeq 
gives rise to an unbroken supersymmetry for every covariantly constant
spinor on $K$ satisfying in addition
\beq
i \gamma_{0123} \epsilon_0 = \epsilon_0.
\eeq
As shown e.g. in \cite{bilal}, this corresponds precisely to picking
out the 16 Poincare supersymmetries of ${\cal N}=4$ SYM and breaks
the 16 special supersymmetries of the superconformal algebra.
In our case, $K$ is flat $R^6$ and these correspond to the 16
Poincare supersymmetries of AdS$_5$. Writing the background spinor
like this, it is obvious that the D7 brane just breaks half
of the Poincare supersymmetries for any warpfactor $Z$,
including AdS$_5$ $\times$ $S^5$, as it does in flat space, since
all projection matrices can be pulled through $Z^{\frac{1}{8}}$.
The same arguments show, that for the D7-D7' system the 
4 supersymmetries that are preserved by the D3-D7-D7' configuration
in the flat embedding space are still preserved once we include
the backreaction of the D3 branes; that is go over to the warped geometry.
In the massless case, in addition we expect to recover 4 more
supersymmetries from the special superconformal supersymmetries.

Moreover, it was shown by Grana and Polchinski in \cite{grana2} that the
same supersymmetries are preserved when we turn on, in addition, 
certain 3-form fluxes on $K$ (the (0,3) piece has to vanish).
This includes in particular the Klebanov Strassler solution \cite{ks},
which we will
investigate in the next section.

\section{Chiral symmetry and the Conifold}
\subsection{Chiral symmetry from NS5 branes}
The problem of how to incorporate chiral flavors in brane setups
was first solved in \cite{brodie} in the context of Hanany-Witten
setups \cite{HW}, which are related to orbifolds of the D3-D7 system
by a simple T-duality \cite{luest, gremm, uranga}.

\begin{figure}
 \centerline{\psfig{figure=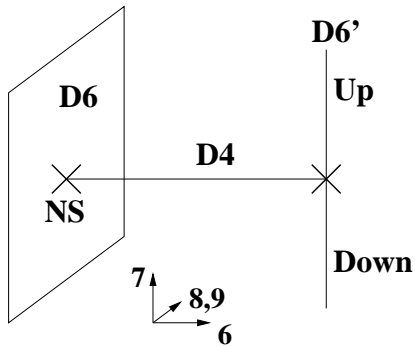,width=2.2in}}
 \caption{${\cal N}=2$ SYM with ${\cal N}=2$ flavors
from D6 branes and ${\cal N}=1$ flavors from D6' branes.}
\label{bro}
  \end{figure}
Starting with a brane setup for an ${\cal N}=2$ gauge theory with
NS5 branes along 012345 and D4 branes along 01236, two types
of D6 branes can be introduced preserving some supersymmetry.
D6 branes along 0123789 add a hypermultiplet worth of matter
with the ${\cal N}=2$ preserving $Q X \tilde{Q}$ superpotential, and
hence no chiral symmetry. The rotated D6' brane along 0123457 switches
off the superpotential, and hence the gauge theory has the full chiral
$SU(M) \times SU(M)$ global symmetry. In the brane setup this is seen
as the fact that the D6' brane can split on the NS5 brane, 
see Fig.(\ref{bro}). The two pieces by
themselves form a HW setup, realizing a 6d $SU(M) \times SU(M)$
gauge theory \cite{ilka}. 

\subsection{Chiral multiplets from D7 branes on orbifolds}
The HW setup we just described has a T-dual description
in terms of D3 branes on an $C^2/Z_k$ orbifold. Following
a suggestion of \cite{gremm}, \cite{uranga} showed that for
an orbifold acting on $u_2$ and $u_3$, the D6 branes map to $u_1=0$
while a single D6' brane maps to two D7' branes,
one D7$_3$ along $u_3=0$ and one D7$_2$ 
along $u_2=0$. A standard orbifold calculation \cite{uranga}
yields that each of those D7' branes only gives rise to a single chiral
multiplet for a single $SU(N)$ gauge group associated with a single
set of
fractional branes. If both types of fractional branes
are present, like in the cascading solution of KS \cite{ks}
for which the supergravity background is known, a single D7'
gives rise to a chiral fundamental in one and a chiral
anti-fundamental in the neighboring gauge group. In more singular
geometries obtained from deformation of $Z_{>2}$ orbifolds, the same
pattern persists: each D7' brane gives rise to two chiral multiplets,
as usual. But they are charged under different gauge groups. If only
a single fractional D3 brane is present, the D7' brane only contributes
a single chiral multiplet. From the Hanany-Witten point of view this
phenomenon was studied in \cite{ilkaami} and was called "flavor-doubling".
On the blown up orbifold,
\beq
xy = \prod_{i=1}^k (z-z_i)
\eeq
the D7' brane wraps the cycle $z=0$, or $xy=\prod z_i$, a single smooth
curve. In the orbifold limit this degenerates into the curve $xy=0$
with the two branches corresponding to the two different D7' branes.
Turning on a mass again corresponds to a vev
for the bifundamental scalar of the D7$_2$ - D7$_3$, deforming the curve
to $xy=\epsilon$, while keeping the $z_i=0$ (as in our
non-chiral toy example).

A new issue that arises in this chiral example is one of anomalies. The number
of chiral and anti-chiral flavors has to be the same in the field theory.
On the gravity side the new ingredient is that a $B$-field is
turned on in the orbifold background. Even though the 7-branes
still wrap a topologically trivial cycle on the base, there now
will be induced lower dimensional D-brane charges, and tadpoles become
an issue. In the orbifold limit a perturbative string theory calculation
\cite{uranga}
shows that tadpoles indeed cancel, if and only if, the number of D7' branes
along $u_2$=0 and $u_3=0$ respectively are the same.

\subsection{The conifold}

In order to get a gravity dual for a confining gauge theory we turn
to the conifold. The corresponding gauge theory is pure ${\cal N}=1$ SYM
if we study a single set of $N$ fractional 3-branes and the corresponding
background was found in \cite{ks} \footnote{More
precisely, the KS solution evolves through a duality cascade
through product gauge groups of the form $SU(N) \times SU(N+M)$.
The physics of the last cascade is (almost) that of pure SYM.}. 
In the T-dual
brane setup \cite{urangaconi,dasgupta}, now in addition to
rotating the D6 brane we also rotate one of the NS5 branes. So this
time, no matter whether we introduce D6 or D6' branes, we will get 
matter with a chiral symmetry
and the corresponding gauge theory is $SU(N)$ SYM with
$M$ flavors. The deformed conifold is given by
\beq
w_1^2 + w_2^2 +w_3^2 +w_4^2 = const.,
\eeq
specifying the internal space in the KS solution.
The D7 brane, T-dual to the D6' brane which is not affected by the rotation 
of the NS5 brane, still lives on the curve
\beq
w_1^2 + w_2^2 = (w_1 + i \, w_2) (w_1 - i \, w_2) = \epsilon,
\eeq
where $\epsilon$ is the mass of the multiplet. For $\epsilon=0$
we obtain two sets of D7 branes, giving rise to the chiral $SU(M) \times
SU(M)$ global symmetry, with the bifundamental scalar being dual to the
superpotential mass term. 
As we recalled above, \cite{grana2} showed that the supersymmetries
of the KS background are once more given by
the Killing spinors of the internal space multiplied by a power
of the warp factor, despite the additional 3-from flux.
As shown in \cite{andy}, $\kappa$-symmetry
again would require that our 7-brane lie on a holomorphic curve inside
the conifold, even in the presence of the B-field. In addition, the
pullback of the B-field has to be (1,1) and we would need to verify
that it satisfies an extra condition on the worldvolume.  Though
we have not explicitly done this, we expect from the T-dual setup that
the supersymmetric cycle is indeed given by the above curve.
In terms of the "usual" coordinates
used to describe the singular, warped conifold, i.e. those used in
\cite{ks}, our zero mass cycle is simply given by $\theta_1=0$ or
$\theta_2=0$, very similar to the $S^5$ case.

\section{Conclusions}

We have shown that a few flavors can be incorporated into
the closed string dual of large $N$ gauge theories by adding
space-time filling probe D-branes. Several unusual properties of
D-branes on AdS spaces have been uncovered, most of them due
to the dynamics associated with the Breitenlohner-Freedman bound.
What remains to be done is a quantitative study of the Klebanov-Strassler
solution, which should be possible with the tools we provided.
In particular, ${\cal N}=1$ SYM has a runaway superpotential, which
should be possible to extract by finding the embedding 
of the probe D7 branes in the deformed conifold geometry.  This 
runaway may be stabilized by adding a mass term, in which case 
the scalar dual to the meson vev will have a profile proportional
to $\frac{\Lambda^3}{m}$.  Alternatively, a quartic potential might be part 
of the field theory dual to the \cite{ks} solution with flavors, and a stable 
vacuum would then exist even for $m=0$. For this we have to study the 
supersymmetry
conditions for D7 branes in the warped conifold background more closely,
we hope to do so in the future.
Another theory for which our tools can be applied is pure ${\cal N}=2$
with ${\cal N}=1$ preserving fundamental matter,
a theory whose dynamics is rather
intricate \cite{kutsei}. Ultimately we believe that our approach will also be
useful for non-supersymmetric large $N$ QCD. Once the supergravity
background is known, including a few flavors corresponds to adding
two distinct sets of space-time filling D-branes.
The chiral condensate corresponds to a nontrivial profile for the
bifundamental scalar. The meson-spectrum can be read off from
the discrete eigenmodes of the worldvolume gauge fields,
just like the glueball masses arise as the eigenmodes of the graviton.
The chiral Lagrangian appears as the effective action of the zero-modes.
Of course, to get $M=N=3$ the backreaction of the D-branes has to
be included. Neglecting the backreaction, we could still
compare to quenched lattice calculations.

\appendix
\section{Image of a Holomorphic Curve on $S^5$}

Again we 
want to use eq.(\ref{emb}) to determine the curve the single
smooth D7 brane wraps in AdS$_5$ $\times$ $S^5$ and establish that
it is a supersymmetric solution. 
This time we parametrize the $S^5$ as
\begin{eqnarray*}
Re(u_1)=x_4 &=& u \; \sin(\psi) \; \cos(\theta) \\ 
Im(u_1)=x_5 &=& u \; \sin(\psi) \; \sin(\theta) \\ 
Re(u_2)=x_6 &=& u \; \cos(\psi) \; \sin(\xi) \; \cos(\alpha) \\ 
Im(u_2)=x_7 &=& u \; \cos(\psi) \; \sin(\xi) \; \sin(\alpha) \\ 
Re(u_3)=x_8 &=& u \; \cos(\psi) \; \cos(\xi) \; \cos(\beta) \\ 
Im(u_3)=x_9 &=& u \; \cos(\psi) \; \cos(\xi) \; \sin(\beta) .
\end{eqnarray*}
The metric is
\beq
ds^2= d \psi^2 + \sin^2 (\psi) d \theta^2 + \cos^2 (\psi) \left (
\sin^2 (\xi) d \alpha^2 + d \xi^2 + \cos^2(\xi) d \beta^2 \right )
\eeq
with $\alpha$, $\beta$ and $\theta$ running from 0 to $ 2 \pi$ and
$\psi$ and $\xi$ running from 0 to $\frac{\pi}{2}$.
The zero mass curve
\beq
u_2 u_3 =0 \; \Leftrightarrow ( x_6=x_7=0 \mbox{ or } x_8=x_9=0 )
\eeq
is just given by two $S^3$s
\beq
\xi=0  \; \mbox{ or } \;\xi=\frac{\pi}{2}
\eeq
intersecting along the $S^1$ $\psi=\frac{\pi}{2}$. Turning on $\epsilon$ which
we choose to be real 
we get
\beq x_6 x_8 - x_7 x_9 = \epsilon \; \mbox{ and }\; x_6 x_9 +
 x_7 x_8 =0
\eeq
which yields
\beq
u^2 \cos (\psi) \sin(2 \xi) \sin (\alpha + \beta) =0 \; \mbox{ and }\;
u^2 \cos (\psi) \sin(2 \xi) \cos (\alpha + \beta) = 2 \epsilon 
\eeq
and hence, taking $\alpha$ and $\xi$ to be the two fluctuating fields we 
find the solution
\beq \alpha=-\beta \; 
\;\mbox{ and } \;\sin(2 \xi) =  \frac{2 \epsilon}{u^2}
 \frac{1}
{\cos (\psi)}. \eeq
As in the D3 D5 case this restricts $u$ to reach a minimal value $u^2_{min}=
2\epsilon$. In addition for a given $u \geq u_{min}$ the $\psi$ variable
only runs over values such that $ \frac{2 \epsilon}{u^2 \cos(\psi)}
 \leq 1$.
At the boundary, $u \rightarrow \infty$, we just recover two equatorial $S^3$s.
From the 5d point of view all that happened is that the bifundamental
scalar from the DD' strings develops a profile, higgsing the $SU(M) \times
SU(M)$ gauge symmetry down to its diagonal subgroup.

\noindent{\large\bf Acknowledgments:}
We would like to thank Mina Aganagic, Ofer Aharony,
Neil Constable, Oliver DeWolfe, Dan Freedman,
Michael Gutperle, Shiraz Minwalla, Lisa Randall,
Stephe Sharpe and Matt Strassler for helpful comments.
This work was partially supported 
by the DOE under contract DE-FGO3-96-ER40956. Further
AK likes to acknowledge the Physics Department of Harvard University
for its support during the initial stages of this work.

\vfill
\eject

\bibliography{flavor}
\bibliographystyle{ssg}
\end{document}